# Chemical-Disorder-Induced Non-metallic Transport in Thermodynamically Metallic $Mo_4TGa_{16}Ge$ (T = Co, Rh or Ir)

Chaoguo Wang,[1] Jiaqi Tian,[1] and Xin Gui[1]*

[1] Department of Chemistry, University of Pittsburgh, Pittsburgh, PA, 15260, USA

*Correspondence to: xig75@pitt.edu

**Abstract**

Flat-band electronic states can be highly sensitive to chemical perturbations, offering opportunities to access a broader range of electronic behaviors beyond that of parent materials. Here, we report the discovery of a new series of compounds, $Mo_4TGa_{16}Ge$ (T = Co, Rh, or Ir), derived from the strongly correlated flat-band superconductor, $Mo_4PtGa_{17}$, *via* nominally preserving the total valence electron counts. All three materials crystallize in a noncentrosymmetric cubic space group, *F*-43*m*, with Ge selectively occupying one of the Ga sites in the parent compound. Although the total electron count remains the same, $Mo_4TGa_{16}Ge$ exhibits significantly distinct properties from $Mo_4PtGa_{17}$. Theoretical calculations predict metallic electronic structures with narrow, dominant Mo-*d* states at the Fermi energy, while low-temperature heat capacity results demonstrate finite Sommerfeld coefficients. However, electrical transport measurements show predominantly non-metallic behaviors with small observed activation energies, excluding the possibilities of conventional semiconducting behaviors. Combining chemical bonding analysis, electronic structure and compositional determination from X-ray crystallography and spectroscopy, we propose that the coexistence of the non-metallic transport features and the thermodynamically metallic behaviors can be attributed to the high sensitivity of the narrow-band materials to chemical disorders. Thus, the discovery and investigation of $Mo_4TGa_{16}Ge$ provide a chemically tunable platform for studying disorder-controlled transport in flat-band intermetallics.

## Introduction

Quantum materials with flat or weakly dispersive electronic bands near the Fermi level ($E_F$) provide fertile platforms for emergent quantum phenomena, such as strong electronic correlation,[1–3] quantum criticality,[4,5] and unconventional superconductivity[6,7] . Due to their small kinetic energies, flat-band-derived states can be highly sensitive to changes in band filling, orbital hybridization, lattice geometry and spin-orbit coupling.[8–10] This makes controlled tuning of flat-band systems an important pathway to manipulating their electronic ground states. Systematic tuning has been conducted on a variety of flat-band materials through chemical substitution,[11,12] pressure,[13,14] strain[15,16] and electrostatic control.[17–21] Recently, an interesting system, $Mo_4PtGa_{17}$, was reported to host a geometrically frustrated breathing pyrochlore lattice of Mo, which produces nearly flat electronic bands and van Hove singularities close to $E_F$.[22] Meanwhile, $Mo_4PtGa_{17}$ was found to be a rare example of a *d*-electron heavy-fermion-like superconductor accompanied by strong electronic correlations and dominant ferromagnetic spin fluctuations, while it is isostructural and isoelectronic with a reported itinerant ferromagnet, $Cr_4PtGa_{17}$.[23] The intriguing combination of quantum states in $Mo_4PtGa_{17}$ was then believed to originate from the Mo-*d*-derived nearly flat electronic bands. These features make $Mo_4PtGa_{17}$ an attractive parent system for exploring how chemical modification of a frustration-derived narrow-band electronic structure can generate new electronic states.

The material design was inspired by two previous studies on the same structural family: $Mo_4FeGa_{17-x}Ge_x$ ($x \sim 2.1$)[24] for which the total valence electron count remains identical to that of $Mo_4PtGa_{17}$, where Fe occupies the Pt site while Ge selectively occupies one of the Ga sites, i.e., Ga' site (Figure 1a), and Sb-doped $Cr_4PtGa_{17}$[25] where Sb also shows site selectivity on the same Ga' site. Together, these observations identify the Ga' site as a chemically addressable site for main-group substitution and suggest a route for expanding the compositional space of this structural family while retaining the total valence electron count. Moreover, such site selectivity is particularly useful for modifying the local chemical environment and electronic structure without introducing extensive chemical disorders across multiple Ga sites.

Here, we report the discovery of the valence-electron-count-preserving series, $Mo_4TGa_{16}Ge$ (T = Co, Rh, Ir), in which Ge selectively occupies the Ga' site and which crystallizes in a noncentrosymmetric cubic space group, $F\bar{4}3m$. Although the total valence electron count is preserved, all three new materials exhibit electronic behavior significantly different from that of the metallic parent compound, $Mo_4PtGa_{17}$. $Mo_4TGa_{16}Ge$ shows low-energy electronic contributions to heat capacity and Pauli-like paramagnetism, consistent with a theoretically predicted finite density of states at $E_F$, while, surprisingly, the electrical transport is predominantly non-metallic. By analyzing chemical bonding, electronic structure and chemical compositions, we propose that chemical disorders can be the reason for such a paradox. The resulting coexistence of metallic thermodynamic signatures with non-metallic transport properties places $Mo_4TGa_{16}Ge$ among an unusual class of materials in which the presence of low-energy electronic states is decoupled from conventional metallic charge transport, such as FeCrAs[26] and $Lu_2Rh_2O_7$[27]. The discovery of $Mo_4TGa_{16}Ge$ therefore expands the chemical and electronic landscape of the $Mo_4PtGa_{17}$ family and provides a platform for examining how local chemical order/disorder controls charge transport in narrow-band intermetallics.

**Results and Discussion**

**Crystal structure and site selectivity of Ge:** $Mo_4TGa_{16}Ge$ (T = Co, Rh, Ir) crystallizes in a noncentrosymmetric cubic space group, $F\bar{4}3m$ (S.G. 216), as shown in Figure 1a. The crystallographic refinement data, atomic sites, and anisotropic thermal displacement results are shown in Table 1 and Tables S1 & S2 in the Supporting Information (SI). $Mo_4TGa_{16}Ge$ consists of one crystallographically equivalent Mo site, which forms $Mo_4$ tetrahedra that are arranged in a face-centered cubic (*fcc*) setting (Figure 1a). Meanwhile, the T atom occupies one atomic site that also adopts a *fcc* sublattice. Disorders are found on the Ga sites in $Mo_4IrGa_{16}Ge$, as shown in Figure S1 in SI. Additionally, three main coordination types are presented in Figure 1b, illustrating a Mo@$Ga_9$ triply capped trigonal prism, a T@$Ga_6$ octahedron and a Ge@$Ga_4$ tetrahedron. Similar to the reported $M_4PtGa_{17}$ (M = Cr or Mo)[22,23], a Mo breathing pyrochlore lattice is also seen in $Mo_4TGa_{16}Ge$ (Figure 1c). However, $M_4PtGa_{17}$ was found to crystallize in a rhombohedral unit cell, $R3m$, a subgroup of $F\bar{4}3m$. The structural relationship between the two space groups is illustrated in Figure 1d,where $Mo_4$ tetrahedra, Pt/T atom, and Ga'/Ge atom demonstrate identical conformations, i.e., the [0001] direction in rhombohedral $M_4PtGa_{17}$ mimics the [111] direction in cubic $Mo_4TGa_{16}Ge$.

Notably, given the chemical similarities between Ga and Ge atoms, Ge is determined by single-crystal X-ray diffraction (XRD) to selectively occupy only one Ga site, i.e., the Ga' site in Figure 1d, whereas no obvious Ge/Ga disorders are observed from single-crystal XRD refinements, i.e., $Mo_4TGa_{16}Ge$ isa chemically ordered phases based on XRD results. However, additional measurements are necessary to accurately determine if Ge/Ga disorders exist. While all the other Ga sites bond with the nearest Mo and T atoms, interestingly, the Ga'(Ge) site does not show such a tendency due to the long interatomic distances, i.e., $d_{Mo\text{-}Ga'(Ge)}$ ~ 5 Å and $d_{T\text{-}Ga'(Ge)}$ ~ 4.7 Å; instead, it forms bonds with four other Ga atoms and results in the Ga'(Ge)@$Ga_4$ tetrahedron shown in Figure 1b. The chemical selectivity was also observed in Sb-doped $Cr_4PtGa_{17}$[25] and $Mo_4FeGa_{17.25\text{-}x}Ge_x$[24] where Sb and Ge atoms preferentially occupy the same Ga'(Ge) site with no clear chemical disorders seen on other Ga sites.

**Phase purity and compositional determination:** As mentioned in the Experimental Section, $Mo_4TGa_{16}Ge$ was synthesized using Ga flux and small crystals were yielded, as can be seen in Figure S2 in SI. However, the obtained crystal size is insufficient to determine orientation-dependent properties. Therefore, small crystals were manually picked for all the following measurements. To test the phase purity of the crystals, powder XRD patterns were collected on the crushed crystals and shown in Figure 2. Rietveld refinement using Fullprof Suite[28] was conducted to confirm the consistency between the observed patterns and the crystal structures determined by single-crystal XRD. The obtained refinement parameters, i.e., $R_p$, $R_{wp}$, and $\chi^2$, indicate high sample homogeneity and high-quality fitting. The obtained lattice parameters are 11.52234 (1) Å (T = Co), 11.57925 (1) Å (T = Rh), and 11.58907 (1) Å (T = Ir), comparable to those from single-crystal XRD, i.e., 11.525 (1) Å (T = Co), 11.554 (1) Å (T = Rh), and 11.587 (1) Å (T = Ir). The discrepancies might originate from the higher temperatures in the sample chamber of the powder X-ray diffractometer (typically ~ 40 – 50 ºC).

To confirm the chemical compositions, the Scanning Electron Microscope with Energy Dispersive X-ray Spectroscopy (SEM-EDS) was performed on $Mo_4TGa_{16}Ge$ crystals. The results shown in Table S3 in SI reveal chemical compositions similar to $Mo_4TGa_{16}Ge$ within the standard deviation, validating the single-crystal-XRD-determined formulas.

**Chemical bonding analysis:** Efforts at synthesizing $Mo_4TGa_{17}$ (T = Co, Rh, Ir), i.e., without Ge, failed to yield the target phases using the same reaction conditions described in the Experimental Section, suggesting that the inclusion of Ge and the selection of T atoms may stabilize these phases, and that maintaining a valence electron count identical to that of the parent compound, $Mo_4PtGa_{17}$, is important. To understand how the identity of Ga'(Ge) and T sites affects chemical bonding and the stability of the resulting phases, Crystal Orbital Hamilton Population (-COHP) and Integrated Crystal Orbital Hamilton Population (-ICOHP) are plotted in Figure 3, which reveal a significant redistribution of chemical bonding upon going from $Mo_4PtGa_{17}$ to isoelectronic $Mo_4TGa_{16}Ge$ derivatives.

As shown in Figures 3a – 3d, Mo-Ga, Ga'-Ga/Ge-Ga, T-Ga and Ga-Ga are dominantly bonding interactions near the Fermi level ($E_F$), while the bonding contributions from Ga'-Ga/Ge-Ga and Mo-Ga are nearly identical in all samples, whereas Mo-Mo shows antibonding interactions around $E_F$. The major distinctions in -COHP are threefold:

*a.* The T-Ga bonding interactions contribute similarly at $E_F$ for $Mo_4PtGa_{17}$ and $Mo_4TGa_{16}Ge$ (T = Rh, Ir), whereas they are negligible in $Mo_4CoGa_{16}Ge$, consistent with stronger hybridization between the more spatially extended 4*d*/5*d* orbitals and Ga-*p* states.

*b.* The Mo-Mo antibonding interaction is much stronger in $Mo_4CoGa_{16}Ge$ compared to the others. The fact that Mo-Mo shows antibonding features across the series suggests that the low-energy electronic structure is governed primarily by metal-Ga hybridization rather than direct Mo-Mo bonding.

*c.* The Mo-Ga and Ga'-Ga/Ge-Ga bonding interactions are stronger than those for T-Ga in $Mo_4PtGa_{17}$ and $Mo_4CoGa_{16}Ge$, while the opposite is found in $Mo_4RhGa_{16}Ge$ and $Mo_4IrGa_{16}Ge$.

Moreover, -ICOHP in Figures 3e – 3h describes the bonding strength and further demonstrate that the changes in -COHP are not limited to the low-energy regime but extend to the overall bonding network. It can be seen that Ga'-Ga/Ge-Ga, Mo-Mo and Ga-Ga interactions are overall bonding in nature above -5 eV, indicating their contribution to stabilizing the phases. Two major distinctions in -ICOHP are:

*a.* The Mo-Ga bonding contribution in $Mo_4PtGa_{17}$ is significantly larger than those in the Co/Rh/Ir compounds.

*b.* The Co-Ga bond makes only a small contribution to the total bonding, while the Rh-Ga, Ir-Ga and Pt-Ga bonds are comparable. When attention is restricted to the $Mo_4TGa_{16}Ge$ series, Mo-Ga dominates the bonding interaction in T = Co, while a more balanced distribution between T-Ga and Mo-Ga can be seen in T = Rh and Ir.

To better understand the -COHP and -ICOHP results, the crystal structure of this system is presented in Figure 3i with all polyhedra shown. Note that although $Mo_4PtGa_{17}$ was reported to crystallize in the space group of $R3m$, different from $Mo_4TGa_{16}Ge$, it adopts a pseudo-cubic symmetry and can be approximately treated as having the same crystal structure as $Mo_4TGa_{16}Ge$. Here, the local polyhedral connectivity provides a possible chemical origin for the observed -COHP/-ICOHP behaviors. The Ga'/Ge@$Ga_4$ tetrahedron and the T@$Ga_6$ octahedron are both connected to the Mo@$Ga_9$ polyhedron by sharing Ga atoms, while the Ga'/Ge@$Ga_4$ tetrahedron and the T@$Ga_6$ octahedron do not directly connect with each other. Consequently, the substitution of Ga' with Ge can perturb one subset of the Ga atoms involved in the Mo-Ga bonding, whereas switching from Co to Rh and to Ir independently modifies another subset through the T-Ga bonding. Because Ge is more electronegative than Ga, the replacement of Ga' by Ge may polarize the electronic density of the shared Ge(Ga')-Ga-Mo network and then modify the energy and hybridization of the Ga-*p* states that simultaneously participate in Mo-Ga bonding. Within this chemical picture, Ge substitution establishes a common modification of the Mo-Ga framework, while the identity of T atoms provides a second tuning parameter that progressively strengthens T-Ga covalency from Co to Ir.

**Pauli-like paramagnetic behavior:** The temperature-dependent magnetic susceptibility of $Mo_4TGa_{16}Ge$, $\chi(T)$, is measured under an external field of 0.5 T under zero-field protocol on manually picked crystals, as shown in Figure 4a. The diamagnetic background from sample containers has been subtracted for all samples. Here, a temperature-independent behavior can be observed in all samples. Susceptibility can generally be considered as the combination of the temperature-independent susceptibility ($\chi_0$), and a temperature-dependent paramagnetic contribution. The temperature-dependent paramagnetic contribution in $Mo_4TGa_{16}Ge$ is considered trace, based on the itinerant magnetic nature of the other reported isostructural phases.[22,23] The temperature-independent $\chi_0$ is thus the dominant contributor of magnetism in $Mo_4TGa_{16}Ge$, which typically consists of Pauli paramagnetism (positive), Van Vleck orbital paramagnetism (positive) and core diamagnetism (negative). $Mo_4TGa_{16}Ge$ displays positive susceptibility at low temperatures and negative susceptibility at high temperatures. This magnetic behavior can be understood as competing interactions between the positive paramagnetic moments and negative core diamagnetism. As the temperature increases, the paramagnetic contribution is gradually suppressed, allowing the negative core diamagnetism to dominate the total susceptibility. This can be supported by the negative fitted $\chi_0$ ($-8.179 \times 10^{-4}$ emu/Oe/mol) in the parent compound, $Mo_4PtGa_{17}$, suggesting a large negative core diamagnetic contribution in this structural framework, which is of the same order of magnitude as the observed total susceptibility in $Mo_4TGa_{16}Ge$. Therefore, it can be concluded that $Mo_4TGa_{16}Ge$ exhibits Pauli paramagnetic behavior with strong core diamagnetism.

This is further consistent with the field-dependent magnetization of $Mo_4TGa_{16}Ge$, as shown in Figure 4b. $Mo_4CoGa_{16}Ge$ exhibits a weak positive magnetization, which increases with a larger external field, consistent with its Pauli paramagnetic behavior. $Mo_4RhGa_{16}Ge$ and $Mo_4IrGa_{16}Ge$ show negative magnetization with a mostly linear dependence on the field, which indicates dominant core diamagnetism in these two samples, consistent with what is observed in $\chi(T)$.

**Metallic thermodynamic signature:** The parent compound, $Mo_4PtGa_{17}$, shows strong electronic correlations.[22] To determine if such a behavior persists when the total electron count stays identical in $Mo_4TGa_{16}Ge$, the temperature-dependent heat capacity, $C_p(T)$, was measured. As shown in Figure 5 and Figure S3, neither a λ-anomaly nor broad peaks were observed between 1.9 K and 40 K, indicating the absence of long-range magnetic/electronic phase transitions or short-range magnetic ordering. The low-temperature heat capacity of $Mo_4TGa_{16}Ge$ was fitted using

$$C_p = C_{el} + C_{ph} + C_{mag}$$

where $C_{el} = \gamma T$, $C_{ph} = \beta_1 T^3 + \beta_2 T^5$ and $C_{mag}$ stand for the electronic, phononic and magnonic contributions, respectively. $C_{mag} = 0$ due to the lack of strong spin-spin correlations determined from magnetic susceptibility results. The fitted values of $\gamma$ (i.e., Sommerfeld coefficient) are $\gamma_{T=Co}$ = 9.7 (8) mJ/mol/K$^2$, $\gamma_{T=Rh}$ = 7.2 (6) mJ/mol/K$^2$ and $\gamma_{T=Ir}$ = 13.4 (6) mJ/mol/K$^2$, while the fitted $\beta_1$ and $\beta_2$ are summarized in Table S4. The obtained values of $\gamma$ suggest non-zero electronic density of states (DOS) of $Mo_4TGa_{16}Ge$ at $E_F$, indicating metallic behavior from a thermodynamical perspective.

Moreover, it is notable that the low-temperature $C_p/T$ is not linear with respect to $T^2$, suggesting the contribution of anharmonic phonons (i.e., non-zero $\beta_2$). However, another possibility is that the non-linear behavior originates from the onset of broad transitions, such as Schottky anomalies and short-range magnetic ordering. Therefore, $C_p/T(T)$ was measured under external magnetic fields. As can be seen in the main panels of Figure 5, by applying magnetic fields up to 9 T, no obvious changes are observed in the low-temperature $C_p/T(T)$ curves, suggesting the absence of Schottky anomalies and short-range magnetic ordering, both of which are expected to be modified by external magnetic fields that ultimately lead to changes in $\gamma$. Thus, $\gamma_{T=Co/Rh/Ir}$ should be considered as intrinsic electronic contributions, indicating finite DOS at $E_F$.

**Electronic structure:** To confirm the non-zero $\gamma_{T=Co/Rh/Ir}$ from heat capacity, density functional theory (DFT) calculations were conducted to obtain the low-energy electronic band structure and DOS with consideration of spin-orbit coupling (SOC) on Mo, Rh and Ir atoms. As can be seen in Figure 6, nearly flat bands can be observed near $E_F$ for all three compounds, spanning from ~ -100 meV to ~ 100 meV (T = Co) and from ~ -200 meV to ~100 meV (T = Rh, Ir), revealing metallic electronic structure. Additionally, van Hove singularities (vHS) can also be seen at L point for all three compounds. Both flat bands and vHS can contribute significantly to the DOS peak slightly below $E_F$, similar to that in the parent compound, $Mo_4PtGa_{17}$,[22] where a narrower band was found around $E_F$. Mo-*d* and Ga-*p* states are dominant near $E_F$, likely suggesting strong Mo-Ga hybridization, consistent with $Mo_4PtGa_{17}$.[22] In contrast, T-*d* and Ge-*p* states are negligible in the low-energy regime except for Co-*d*, which contributes significantly below ~ -600 meV. Moreover, since the calculations for $Mo_4TGa_{16}Ge$ were performed on the conventional *F*-centered cubic lattice with Z = 4, the obtained DOS at $E_F$ (i.e., $N(E_F)$) can be normalized to ~ 5.6 states/eV (T = Co), ~ 4.5 states/eV (T = Rh) and ~ 4.0 states/eV (T = Ir), which correspond to ~ 13.2 mJ/mol/$K^2$, ~ 10.6 mJ/mol/$K^2$ and ~ 9.4 mJ/mol/$K^2$, respectively. The DFT-derived Sommerfeld coefficients are of the same magnitude as the experimental ones, while the discrepancies between the two can be explained by the existence of chemical disorder, as described below.

Additionally, the comparison between -COHP and DFT results can provide a chemical understanding of the near-$E_F$ states. Although the dominant Mo-*d* states suggest that they originate primarily from the Mo breathing pyrochlore lattice, the -COHP analysis instead shows substantial Mo-Mo antibonding character near $E_F$, while outstanding Mo-Ga bonding interactions exist. Therefore, the Mo-*d*-derived bands at $E_F$ should be treated as strongly hybridized states that are stabilized by surrounding Ga frameworks rather than through direct Mo-Mo bonding. The absence of a bandgap at $E_F$ is also consistent with the -COHP results, suggesting metallic electronic structures.

The systematic evolution across the Co, Rh and Ir compounds also correlates with the -COHP results. The calculated N($E_F$) decreases from Co to Ir, while the -COHP and -ICOHP analyses show gradually strengthened T-Ga bonding. This inverse trend may indicate that the increasing T-Ga hybridization redistributes the electronic states away from $E_F$, thereby reducing the N($E_F$). Meanwhile, the substantially weaker Mo-Ga -ICOHP values of all three Ge-containing compounds relative to $Mo_4PtGa_{17}$ indicate that the electronic differences are not only determined by the identity of T, but also the Mo-Ga bonding framework. To summarize, the DFT-derived band structure and DOS, and -COHP/-ICOHP analyses describe $Mo_4TGa_{16}Ge$ as an electronically metallic systems with finite, predominantly Mo-$d$-derived states at $E_F$, while metal-Ga bonding also contributes significantly in the low-energy regime.

**Non-metallic electronic transport:** To confirm the metallic behavior of $Mo_4TGa_{16}Ge$ obtained from thermodynamic measurements and theoretical calculations, electronic transport measurements were conducted. Figures 7a – 7c present the temperature-dependent electrical resistivity, $\rho(T)$, on multidomain crystals similar to what is shown in Figure S2 in the SI. Surprisingly, semiconducting behaviors are observed in all three compounds in which increasing resistivity is seen with decreasing temperatures, inconsistent with the heat capacity results and theoretical analysis. To determine if the semiconducting behaviors are intrinsic, the transport activation energies are fitted in the high-temperature regions where the negative slope of ρ(T) first emerges. The fitting was conducted by employing the Arrhenius equation

$$\rho(T) = \rho_0 e^{\frac{E_a}{k_B T}}$$

where $\rho_0$ is a pre-factor for fitting and is a constant, $k_B$ is the Boltzmann constant, and $E_a$ is the thermal activation energy. Note that the semiconducting bandgap, $E_g$, can be taken approximately as $E_g = 2E_a$. The insets of Figures 7a – 7c show the linear fitting of lnρ(1/T), revealing $E_g$(T = Co) ~ 5.94 meV, $E_g$(T = Rh) ~ 30.76 meV and $E_g$(T = Ir) ~ 12.57 meV. Combining the metallic thermodynamic and theoretical results, the small activation energies, or bandgaps, likely do not correspond to intrinsic semiconducting bandgaps, suggesting that the non-metallic transport behaviors in $Mo_4TGa_{16}Ge$ do not arise from a conventional semiconducting nature.

**Discussion of the paradox between metallic thermodynamic/theoretical behavior and non-metallic transport:** The electronic and transport properties of $Mo_4TGa_{16}Ge$ reveal an apparent discrepancy between the presence of low-energy electronic states at $E_F$ and the inability to effectively conduct charges. Electronic-structure calculations on the chemically ordered structures determined by single-crystal XRD suggest a finite $N(E_F)$ for all three compounds that are dominated by the narrow Mo-$d$ states, while the low-temperature heat capacity results consistently indicate appreciable low-energy electronic contributions. These observations are in sharp contrast to the predominantly non-metallic electrical transport. Moreover, the relatively small $E_g$ and $E_a$ extracted from high-temperature $\rho(T)$, together with the absence of a calculated bandgap at $E_F$, argue against a conventional intrinsic semiconductor picture in which transport is governed by thermal excitation across a true bandgap. This interpretation is further supported by $Mo_4IrGa_{16}Ge$, which exhibits metallic-like electrical transport above ~ 200 K prior to a metallic-non-metallic crossover upon cooling, as shown in Figure 7d. This indicates that the resistivity of $Mo_4IrGa_{16}Ge$ cannot be described by a simple semiconducting model over the entire measured temperature range.

Chemical disorders provide a plausible origin for the observed discrepancies. Single-crystal XRD cannot reliably distinguish between Ga and Ge and determine Ga/Ge disorders due to similar X-ray scattering factors. However, EDS results shown in Table S3 in the SI, which are normalized to the contents of T, yield Ge concentrations slightly above the ideal value of 1 for T = Co and Rh, while they are slightly smaller than 1 for T = Ir. While the small deviations do not by themselves establish Ga/Ge chemical disorders and should be considered within the uncertainty of EDS quantification, they are consistent with the slight compositional disorder or off-stoichiometry that may not be resolved by X-ray crystallography. In fact, due to the chemical similarity of Ga and Ge, it is unlikely for Ge to occupy the Mo site or T site; thus, it is highly likely that Ge exists on other Ga sites other than the Ga'/Ge site for T = Co and Rh, while Ge occupancy on the Ga'/Ge site in T = Ir can be less than unity.

Such chemical disorders can strongly affect the narrow electronic states near $E_F$. Although these states are primarily arising from Mo-$d$, their dispersions are strongly influenced by Mo-Ga, T-Ga and Ga'-Ga/Ge-Ga bonds, as established in the -COHP/-ICOHP section. As a result, variations in Ga/Ge occupancies can introduce spatial fluctuations in the local potential, which may then interfere with the Mo-Ga/T-Ga/Ge-Ga-mediated electron hopping and lead to non-metallic electrical transport. This is a similar scenario to Anderson localization,[29] where a finite low-energy DOS can remain and contribute to the electronic heat capacity, while its reduced spatial coherence or localization strongly suppresses the charge transport. The small activation energies would then represent the effective transport energies associated with localized states and/or proximity to a mobility edge rather than intrinsic bandgaps. Moreover, due to the relatively flat electronic bands near $E_F$ that are derived from a geometrically frustrated breathing pyrochlore lattice, weak chemical disorder may produce disproportionately large changes in their spatial coherence and transport properties, as suggested by tight-binding models.[30] Therefore, the combined structural, compositional, thermodynamic, transport, and theoretical electronic structures/chemical bonding analyses suggest that the unusual coexistence of thermodynamically metallic behavior and nonmetallic transport in $Mo_4TGa_{16}Ge$ may arise from the pronounced sensitivity of its narrow, hybridized Mo-*d* states to chemical disorders.

## Conclusion

In summary, the $Mo_4TGa_{16}Ge$ (T = Co, Rh or Ir) series displays significantly different properties compared to the parent correlated superconducting $Mo_4PtGa_{17}$, although the total valence electron count was retained. The electronic structure calculations are in agreement with heat capacity measurements, suggesting metallic behavior, which is, however, in contrast to the non-metallic electrical transport results. The small observed activation energies and the high-temperature metallic-like transport observed in T = Ir further argue against a simple intrinsic semiconductor picture. Instead, the combined structural, compositional, thermodynamic, transport and theoretical results suggest that the narrow Mo-derived states are unusually sensitive to the local Ga/Ge environment. Because these states are strongly influenced by Mo-Ga/T-Ga/Ge-Ga hybridization/bonding, even a low degree of chemical disorder can substantially perturb the effective electronic hopping and suppress charge mobility without necessarily eliminating the electronic states near $E_F$, consistent with the known sensitivity of frustrated flat-band-derived states to weak disorder. These results demonstrate that preserving nominal electron count and an ordered substitution motif does not guarantee preservation of electronic itinerancy, and establish local chemical order as an important control knob for charge transport in the narrow-band intermetallic systems.

## Experimental Section

**Synthesis of $Mo_4TGa_{16}Ge$ crystals:** $Mo_4CoGa_{16}Ge$ and $Mo_4RhGa_{16}Ge$ were synthesized by mixing molybdenum powder (Thermo Scientific, ~100 mesh, 99.95%), cobalt powder (Thermo Scientific, ~60 mesh, 99.5%) or rhodium powder (Alfa Aesar, ~325 mesh, 99.95%), gallium ingots (Thermo Scientific, 99.9%), and germanium powder (Thermo Scientific, −100 mesh, 99.999%) in alumina crucibles in a molar ratio of 4:1:50:1, with Ga serving as a self-flux. The crucibles were then sealed in evacuated quartz tubes after being purged with argon. Mo4IrGa16Ge was synthesized using the same molybdenum powder, gallium ingots, and germanium powder, together with iridium powder (Alfa Aesar, ~22 mesh, 99.99%), in alumina crucibles in a molar ratio of 4:1:100:1, also using Ga as the flux. The crucibles were similarly sealed in evacuated quartz tubes. The samples were first heated to 1100 °C at a rate of 180 °C/h and held at this temperature for 2 days. They were then cooled to 920 °C at a rate of 180 °C/h and held for 4 h. Subsequently, the temperature was slowly decreased to 820 °C at a rate of 1 °C/h without an additional holding step. Finally, the samples were cooled to 400 °C at a rate of 5 °C/h where the quartz tubes were rapidly inverted and centrifuged to remove the excess Ga flux. The resulting samples were immersed in 1 M HCl to remove residual Ga attached on the surface. The obtained $Mo_4TGa_{16}Ge$ crystals exhibited a triangular morphology with a metallic luster. SEM images of all three samples are shown in Figure S2.

**Single-crystal X-ray diffraction and powder X-ray diffraction.** Single-crystal X-ray diffraction (XRD) measurements were performed at room temperature using a Bruker D8 QUEST ECO diffractometer equipped with APEX5 software and Mo $K_\alpha$ radiation ($\lambda$ = 0.71037A). The crystals were immersed in glycerol, selected, and mounted on a Kapton loop. Single-crystal diffraction data were collected using Bruker SMART software, and corrections for Lorentz and polarization effects were applied. Numerical absorption corrections based on crystal-face indexing were performed using XPREP. The crystal structures were solved by direct methods and refined by full-matrix least-squares refinement on $F^2$ using the SHELXTL package.[31,32] Subsequently, well-crystallized $Mo_4TGa_{16}Ge$ crystals were selected and ground into powder to examine the phase purity by powder X-ray diffraction (PXRD). PXRD measurements were performed using a Bruker D2 PHASER equipped with Cu $K_\alpha$ radiation and a LynxEye-XE detector. Rietveld refinements of the resulting powder diffraction patterns were performed using the FullProf Suite.[28] The calculated patterns were generated based on the crystal structures determined by SCXRD and fitted to the observed diffraction patterns.

**Scanning electron microscopy with energy dispersive X-ray spectroscopy (SEM-EDS).** Chemical composition analysis was performed by scanning electron microscopy (SEM) with energy dispersive X-ray spectroscopy (EDS). A Ziess Sigma 500 VP SEM with Oxford Aztec X-EDS was used with an electron beam energy of 20 kV to capture the crystal image of $Mo_4TGa_{16}Ge$ samples.

**Physical property measurements.** A Quantum Design Physical Property Measurement System (PPMS) equipped with the ACMS II option was employed to measure the DC magnetization from 1.9 to 380 K under a magnetic field of 0.5 T. Field-dependent magnetization measurements were performed from 0 to 9 T at 2 K. Temperature-dependent magnetic susceptibility measurements of $Mo_4TGa_{16}Ge$ were carried out from 1.9 to 380 K under a magnetic field of 0.5 T using a zero-field-cooled (ZFC) protocol. Electrical resistivity measurements were performed using the same instrument with a standard four-probe method over the temperature range of 2–380 K. Platinum wires were attached to the samples using silver epoxy to ensure ohmic contacts. Heat capacity measurements were performed using the standard relaxation method in the PPMS. All physical property measurements were performed on selected as-synthesized crystals or naturally co-grown clusters of crystals.

**Tight-Binding, Linear Muffin-Tin Orbital-Atomic Spheres Approximation (TB-LMTO-ASA):** Tight-Binding, Linear Muffin-Tin Orbital-Atomic Spheres Approximation (TB-LMTO-ASA) using the Stuttgart code was applied to calculate Crystal Orbital Hamiltonian Population (-COHP) and integrated -COHP (-ICOHP) curves which are able to identify bonding/antibonding interaction for compounds.[33–35] All integrated values were generated with a convergence criterion of 0.05 meV and a mesh of 512 *k* points.[36] In the ASA method, overlapping Wigner-Seitz (WS) spheres were employed to filled the space where the symmetry of the potential is treated as spherical with a combined correction on the overlapping part. The WS radii are: 2.89 Å for Mo; 2.78 Å for Pt; 2.56 Å for Co; 2.70 Å for Rh; 2.80 Å for Ir; 2.62 Å for Ga; and 2.69 Å for Ge. Empty spheres are required for the calculation, and the overlap of WS spheres is limited to no larger than 16%.

**Electronic Structure Calculations***:* The electronic band structure and density of states (DOS) of $Mo_4TGa_{16}Ge$ were calculated using the WIEN2k code, which has the full-potential linearized augmented plane wave method (FP-LAPW) with local orbitals implemented.[37,38] The structural lattice parameters obtained from experiments are used for the calculation. For treatment of the electron correlation within the generalized gradient approximation, the electron exchange-correlation potential was used with the parametrization by Perdew et. al. (i.e. the LDA).[39] The conjugate gradient algorithm was applied and the energy cutoff was 500 eV. Reciprocal space integrations were completed over a 8×8×8 Monkhorst-Pack *k*-points mesh.[40] With these settings, the calculated total energy converged to less than 0.1 meV per atom.

## Acknowledgements

This research is supported by the startup funds from the University of Pittsburgh. SEM-EDS performed in the University of Pittsburgh Nanofabrication and Characterization Core Facility (RRID:SCR_05124) and services and instruments used in this project were graciously supported, in part, by the University of Pittsburgh.

## Competing Interests

The authors declare no competing interests.

## Supplementary Information

Atomic sites and thermal displacement parameters from X-ray crystallography; SEM-EDS

quantitative analysis results; phononic contribution coefficients from heat capacity; SEM pictures of crystals; temperature-dependent heat capacity.

**Table 1.** Single crystal structure refinement for $Mo_4CoGa_{16}Ge$ at 293 (2) K, $Mo_4RhGa_{16}Ge$ at 293 (2) K and $Mo_4IrGa_{16}Ge$ at 303 (2) K.

| **Refined Formula** | **$Mo_4CoGa_{16}Ge$** | **$Mo_4RhGa_{16}Ge$** | **$Mo_4IrGa_{16}Ge$** |
|---|---|---|---|
| Temperature (K) | 293 (2) | 293 (2) | 303 (2) |
| F.W. (g/mol) | 1630.80 | 1674.78 | 1764.07 |
| Space group; Z | *F*-43*m;* 4 | *F*-43*m;* 4 | *F*-43*m;* 4 |
| $a$(Å) | 11.5254 (13) | 11.5537 (13) | 11.572 (7) |
| V (Å$^3$) | 1531.0 (5) | 1542.3 (5) | 1550 (3) |
| θ range (°) | 3.061-32.999 | 3.054-33.086 | 3.049-33.027 |
| No. reflections; $R_{int}$ | 9460; 0.0723 | 7801; 0.1103 | 3948; 0.0486 |
| No. independent reflections | 337 | 342 | 339 |
| No. parameters | 19 | 19 | 22 |
| $R_1$: $\omega R_2$ ($I$>2δ($I$)) | 0.0204; 0.0509 | 0.0219; 0.0470 | 0.0182; 0.0415 |
| Goodness of fit | 1.172 | 1.075 | 1.116 |
| Diffraction peak and hole (e$^-$/ Å$^3$) | 1.373; -1.006 | 1.085; -1.304 | 0.844; -1.018 |
| Absolute structure parameter | 0.03 (4) | 0.01 (4) | 0.07 (2) |

**Figure 1 Crystal structure of $Mo_4TGa_{16}Ge$. a.** The crystal structure of $Mo_4TGa_{16}Ge$ (T = Co, Rh, Ir) with coordination environment of Ge shown where orange, dark purple, light blue and grey spheres represent Mo, T, Ga and Ge atoms, respectively. **b.** The coordination environment of Mo, T and Ge atoms, including Mo@$Ga_9$ triply-capped trigonal prism, T@$Ga_6$ octahedron, and Ge@$Ga_4$ tetrahedron. **c.** Breathing pyrochlore sublattice of Mo. **d.** The comparison between the crystal structure of $Mo_4PtGa_{17}$ and $Mo_4TGa_{16}Ge$ (T = Co, Rh, Ir) with different space groups. Only essential atomic sites are shown for clarity.

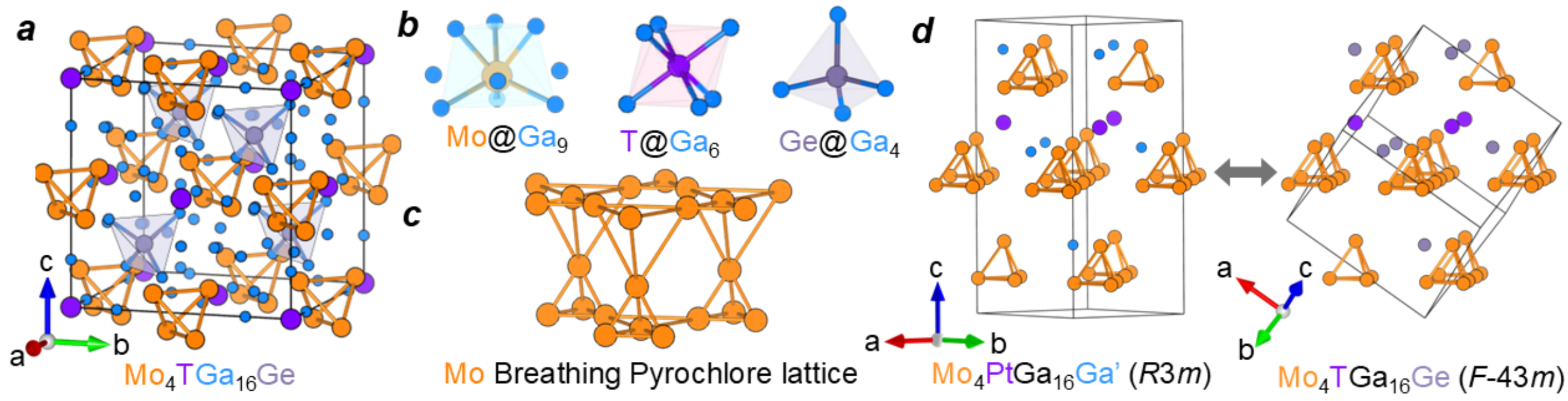

**Figure 2. Phase purity of $Mo_4TGa_{16}Ge$.** Powder XRD patterns of **a.** $Mo_4CoGa_{16}Ge$, **b.** $Mo_4RhGa_{16}Ge$ and **c.** $Mo_4IrGa_{16}Ge$ with Rietveld fitting in red.

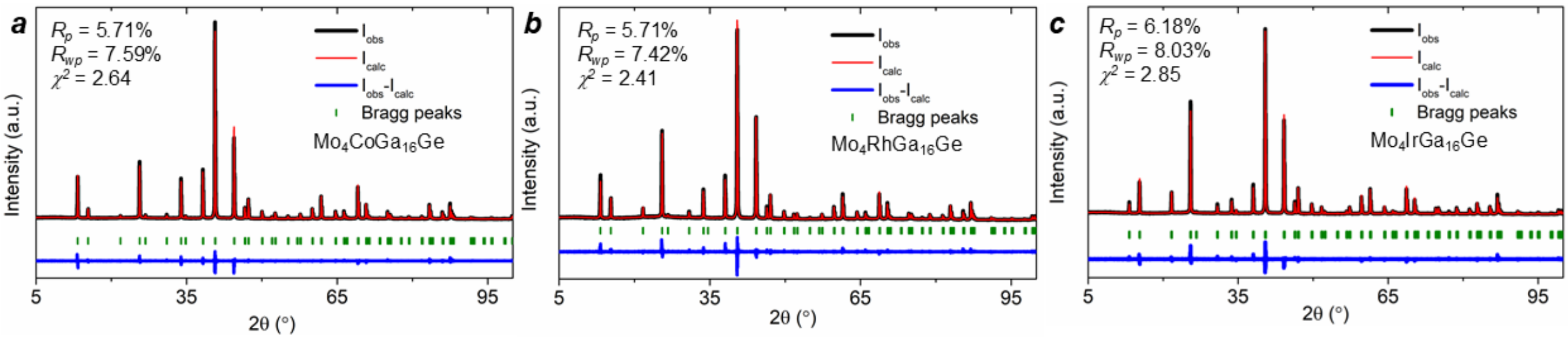

**Figure 3**. **Chemical bonding analysis.** The Crystal Orbital Hamilton Population (-COHP) (**a** - **d**) and Integrated Crystal Orbital Hamilton Population (-ICOHP) (**e** - **h**) demonstrating T-Ga/Pt-Ga, Ga-Ga, Ge-Ga/Ga'-Ga, Mo-Ga and Mo-Mo bonding interactions for **a. & e.** $Mo_4PtGa_{17}$, **b. & f.** $Mo_4CoGa_{16}Ge$, **c. & g.** $Mo_4RhGa_{16}Ge$ and **d. & h.** $Mo_4IrGa_{16}Ge$. The Fermi level is marked by black dash line. **j.** The crystal structure and polyhedral connectivity of $Mo_4PtGa_{17}$ or $Mo_4TGa_{16}Ge$. Note that $Mo_4PtGa_{17}$ exhibits a pseudo-cubic symmetry and can be converted to a similar cubic unit cell as shown. The calculation was performed on the rhombohedral symmetry, instead.

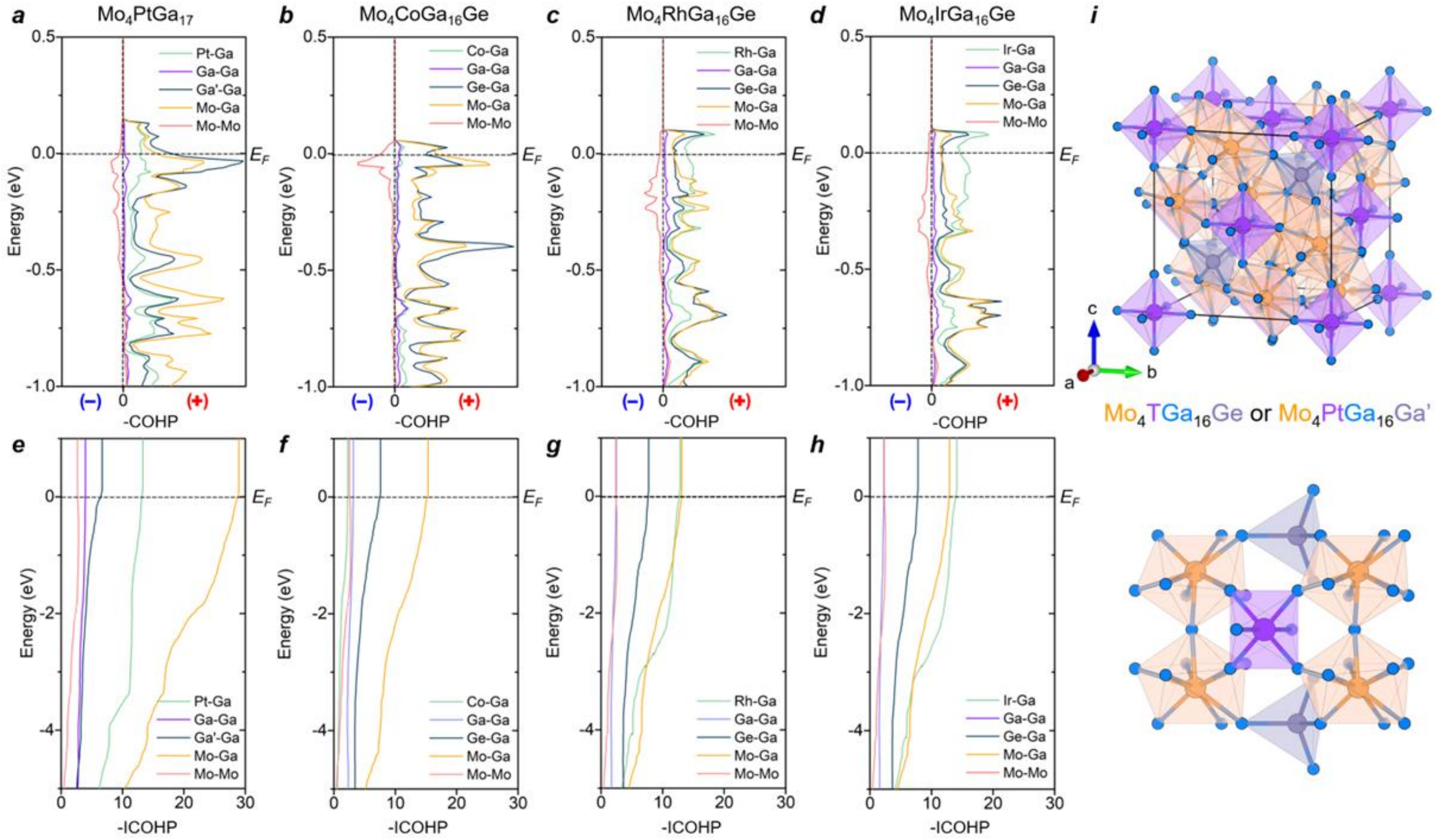

**Figure 4. Pauli-like magnetic behavior. a.** Temperature-dependent magnetic susceptibility ($\chi$) for $Mo_4TGa_{16}Ge$ measured from 2 K to 300 K under an external magnetic field of 0.5 T with sample holder's background subtracted. **b.** Isothermal magnetization under 2 K measured from 0 T to 9 T for $Mo_4TGa_{16}Ge$.

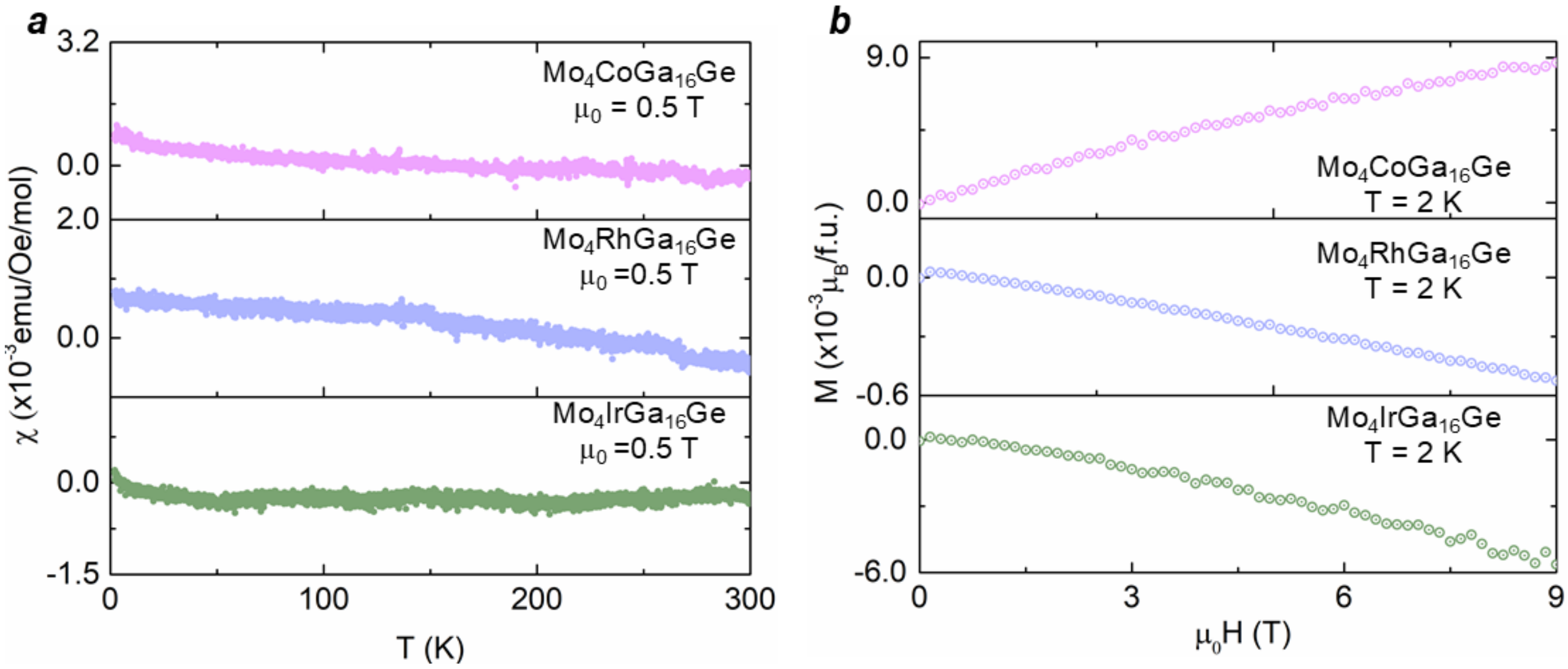

**Figure 5. Metallic thermodynamic signature.** Temperature-dependent heat capacity ($C_p$) measured under 0 T, 5 T and 9 T. **(Main panel)** $C_p/T$ vs T plotted from 0 K to 12 K for **a.** $Mo_4CoGa_{16}Ge$, **b.** $Mo_4RhGa_{16}Ge$ and **c.** $Mo_4IrGa_{16}Ge$. **(Inset).** $C_p/T$ vs $T^2$ with electronic and phononic contributions fitted under low temperature.

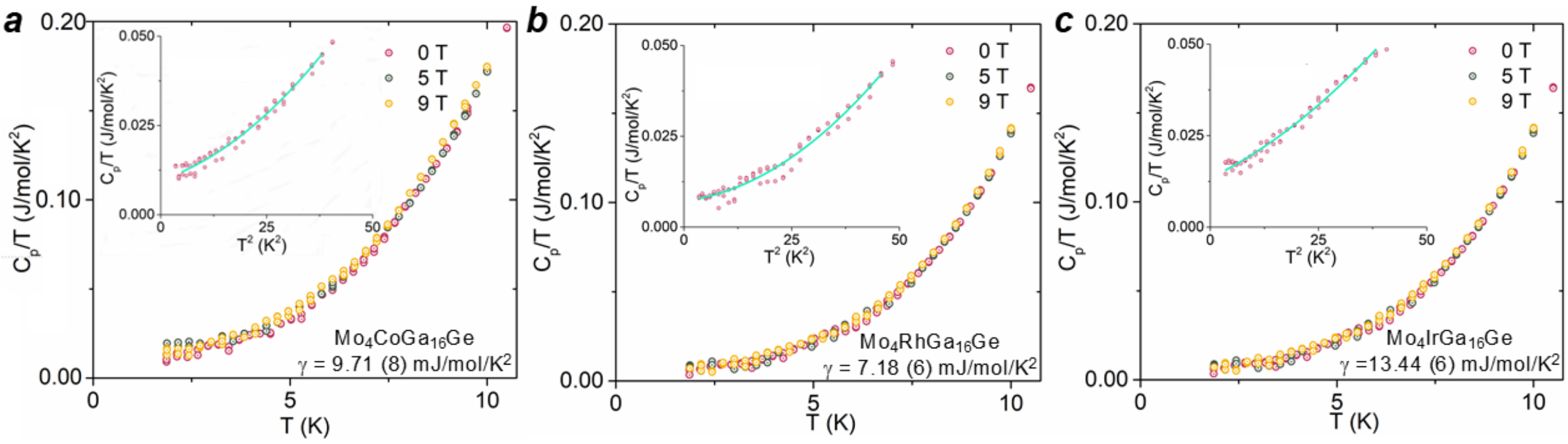

**Figure 6. Electronic structure.** The density functional theory calculations to obtain the low-energy electronic band structure and density of state (DOS) with consideration of spin-orbit coupling (SOC) on Mo, Rh and Ir atoms for **a.** $Mo_4CoGa_{16}Ge$. **b.** $Mo_4RhGa_{16}Ge$ and **c.** $Mo_4IrGa_{16}Ge$.

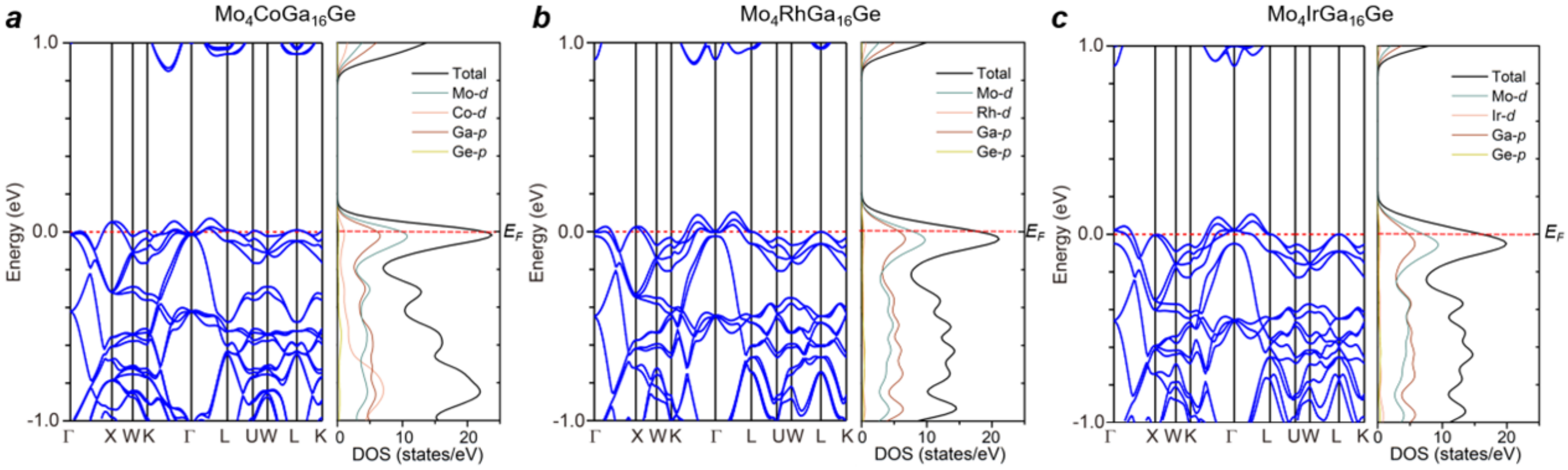

**Figure 7. Electronic transport. (Main panel)** Temperature-dependent electrical resistivity (ρ) under zero magnetic field from 2 to 380 K for **a.** $Mo_4CoGa_{16}Ge$, **b.** $Mo_4RhGa_{16}Ge$ and **c.** $Mo_4IrGa_{16}Ge$. **(Inset)** lnρ vs. 1/T plotted and fitted by Arrhenius equation under linear region to calculate the thermal activation energy. **d.** The enlarged view of ρ(T) for $Mo_4IrGa_{16}Ge$ from 180 K to 280 K with a metal-insulator-like transition.

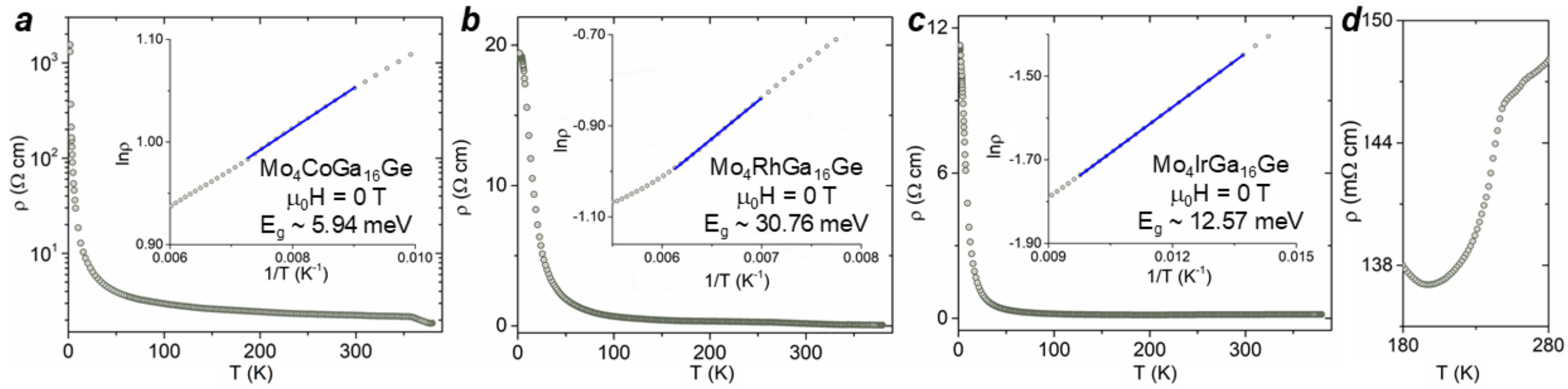

*Supplementary Information*

# Chemical-Disorder-Induced Non-metallic Transport in Thermodynamically Metallic $Mo_4TGa_{16}Ge$ (T = Co, Rh or Ir)

Chaoguo Wang,[1] Jiaqi Tian,[1] and Xin Gui[1]*

[1] Department of Chemistry, University of Pittsburgh, Pittsburgh, PA, 15260, USA

*Correspondence to: xig75@pitt.edu

**Contents**

**Table S1.** Atomic coordinates and equivalent isotropic displacement parameters for $Mo_4TGa_{16}Ge$ at 293 (2) and 303 (2) K. ($U_{eq}$ is defined as one-third of the trace of the orthogonalized $U_{ij}$ tensor (Å$^2$))

**$Mo_4CoGa_{16}Ge$ at 293 (2) K**

| Atom | Wyck. | Occ. | *x* | *y* | *z* | $U_{eq}$ |
|---|---|---|---|---|---|---|
| Co1 | 4*a* | 1 | 0 | 0 | 0 | 0.0066(8) |
| Mo1 | 16*e* | 1 | 0.4041(2) | 0.4041(2) | 0.4041(2) | 0.0045(2) |
| Ge1 | 4*d* | 1 | 0.7500 | 0.7500 | 0.7500 | 0.0069(6) |
| Ga1 | 24*f* | 1 | 0.7959(1) | 0 | 0 | 0.0171(4) |
| Ga2 | 16*e* | 1 | 0.6292(2) | 0.6292(2) | 0.6292(2) | 0.0072(3) |
| Ga3 | 24*g* | 1 | 0.4324(1) | 0.7500 | 0.7500 | 0.0159(4) |

**$Mo_4RhGa_{16}Ge$ at 293 (2) K**

| Atom | Wyck. | Occ. | *x* | *y* | *z* | $U_{eq}$ |
|---|---|---|---|---|---|---|
| Rh1 | 4*a* | 1 | 0 | 0 | 0 | 0.0072(4) |
| Mo1 | 16*e* | 1 | 0.4045(2) | 0.4045(2) | 0.4045(2) | 0.0049(2) |
| Ge1 | 4*d* | 1 | 0.7500 | 0.7500 | 0.7500 | 0.0073(6) |
| Ga1 | 24*f* | 1 | 0.7921(1) | 0 | 0 | 0.0140(3) |
| Ga2 | 16*e* | 1 | 0.6289(2) | 0.6289(2) | 0.6289(2) | 0.0076(3) |
| Ga3 | 24*g* | 1 | 0.4315(1) | 0.7500 | 0.7500 | 0.0130(3) |

**$Mo_4IrGa_{16}Ge$ at 303 (2) K**

| Atom | Wyck. | Occ. | *x* | *y* | *z* | $U_{eq}$ |
|---|---|---|---|---|---|---|
| Ir1 | 4*a* | 1 | 0 | 0 | 0 | 0.0095(2) |
| Mo1 | 16*e* | 1 | 0.4041(2) | 0.4041(2) | 0.4041(2) | 0.0047(2) |
| Ge1 | 4*d* | 1 | 0.7500 | 0.7500 | 0.7500 | 0.0072(6) |
| Ga1 | 24*f* | 1 | 0.7909(1) | 0 | 0 | 0.0197(4) |
| Ga2 | 16*e* | 1 | 0.6294(2) | 0.6294(2) | 0.6294(2) | 0.0083(3) |
| Ga3 | 24*g* | 0.921(5) | 0.4319(2) | 0.7500 | 0.7500 | 0.0134(4) |
| Ga4 | 48*h* | 0.039(3) | 0.4350(1) | 0.6718(17) | 0.8282(17) | 0.0134(4) |

**Table S2.** Anisotropic thermal displacement parameters for $Mo_4TGa_{16}Ge$.

**$Mo_4CoGa_{16}Ge$ at 293 (2) K:**

| Atom | U11 | U22 | U33 | U12 | U13 | U23 |
|---|---|---|---|---|---|---|
| Co1 | 0.0066(8) | 0.0066(8) | 0.0066(8) | 0 | 0 | 0 |
| Mo1 | 0.0045(2) | 0.0045(2) | 0.0045(2) | -0.0001(2) | -0.0001(2) | -0.0001(2) |
| Ge1 | 0.0069(6) | 0.0069(6) | 0.0069(6) | 0 | 0 | 0 |
| Ga1 | 0.0071(7) | 0.0222(6) | 0.0222(6) | 0 | 0 | -0.0065(6) |
| Ga2 | 0.0072(3) | 0.0072(3) | 0.0072(3) | -0.0013(3) | -0.0013(3) | -0.0013(3) |
| Ga3 | 0.0138(7) | 0.0169(5) | 0.0169(5) | 0 | 0 | -0.0105(6) |

**$Mo_4RhGa_{16}Ge$ at 293 (2) K:**

| Atom | U11 | U22 | U33 | U12 | U13 | U23 |
|---|---|---|---|---|---|---|
| Rh1 | 0.0072(4) | 0.0072(4) | 0.0072(4) | 0 | 0 | 0 |
| Mo1 | 0.0049(2) | 0.0049(2) | 0.0049(2) | -0.0003(2) | -0.0003(2) | -0.0003(2) |
| Ge1 | 0.0073(6) | 0.0073(6) | 0.0073(6) | 0 | 0 | 0 |
| Ga1 | 0.0069(6) | 0.0175(5) | 0.0175(5) | 0 | 0 | -0.0025(6) |
| Ga2 | 0.0076(3) | 0.0076(3) | 0.0076(3) | -0.0014(3) | -0.0014(3) | -0.0014(3) |
| Ga3 | 0.0129(6) | 0.0130(4) | 0.0130(4) | 0 | 0 | -0.0071(5) |

**$Mo_4RhGa_{16}Ge$ at 303 (2) K:**

| Atom | U11 | U22 | U33 | U12 | U13 | U23 |
|---|---|---|---|---|---|---|
| Rh1 | 0.0095(2) | 0.0095(2) | 0.0095(2) | 0 | 0 | 0 |
| Mo1 | 0.0047(2) | 0.0047(2) | 0.0047(2) | 0.0000(2) | 0.0000 (2) | 0.0000 (2) |
| Ge1 | 0.0072(6) | 0.0072(6) | 0.0072(6) | 0 | 0 | 0 |
| Ga1 | 0.0074(6) | 0.0259(5) | 0.0259(5) | 0 | 0 | -0.0080(7) |
| Ga2 | 0.0083(3) | 0.0083(3) | 0.0083(3) | -0.0011(3) | -0.0011(3) | -0.0011(3) |
| Ga3 | 0.0118(7) | 0.0142(5) | 0.0142(5) | 0 | 0 | -0.0103(6) |
| Ga4 | 0.0118(7) | 0.0142(5) | 0.0142(5) | 0 | 0 | -0.0103(6) |

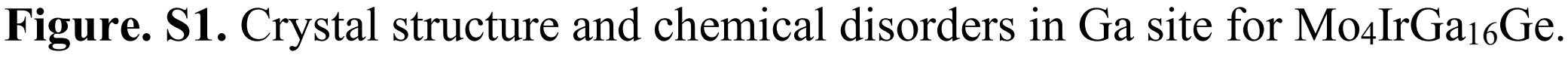

**Figure. S1.** Crystal structure and chemical disorders in Ga site for $Mo_4IrGa_{16}Ge$.

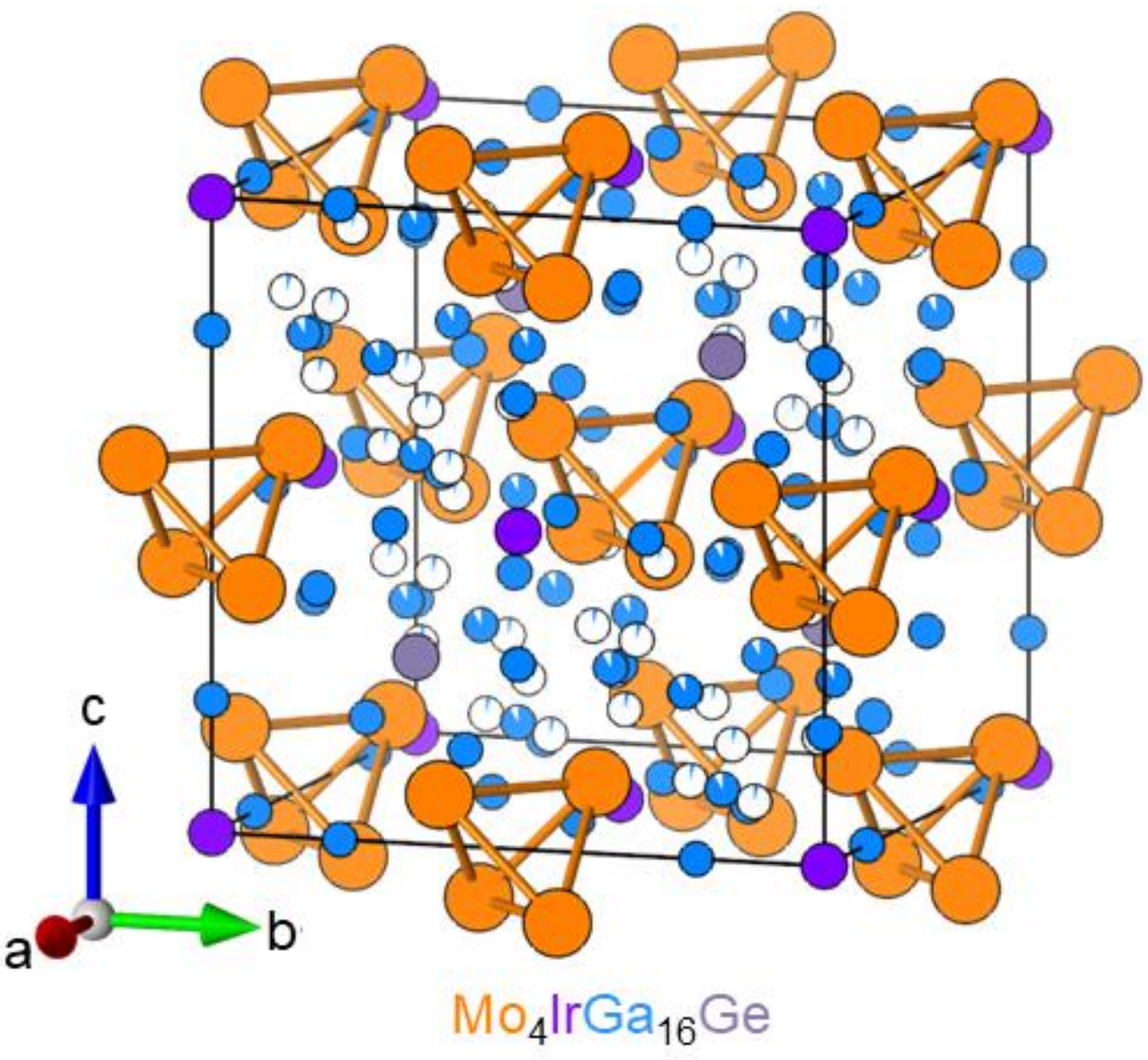

**Figure S2.** The SEM images for small crystals yeild by Ga flux. **a.** $Mo_4CoGa_{16}Ge$. b. $Mo_4RhGa_{16}Ge$ and c. $Mo_4IrGa_{16}Ge$.

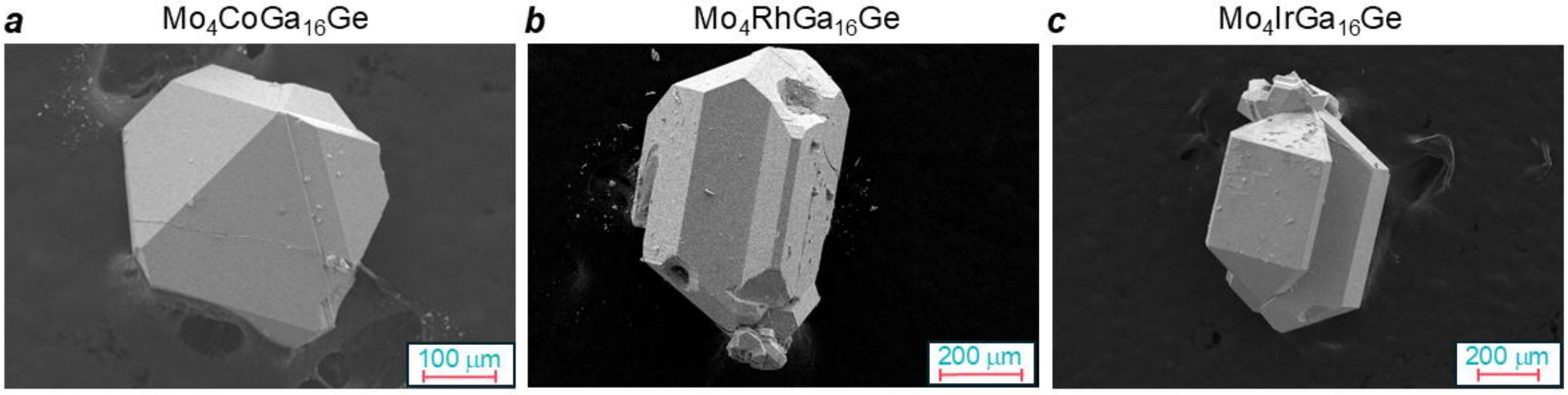

**Table S3.** EDS results for $Mo_4TGa_{16}Ge$.

**$Mo_4CoGa_{16}Ge$**

| Spectrum # | Mo | Co | Ga | Ge |
|---|---|---|---|---|
| **Spectrum 1** | 18.81 | 4.56 | 71.19 | 5.43 |
| **Spectrum 2** | 17.97 | 4.61 | 72.34 | 5.09 |
| **Spectrum 3** | 17.30 | 4.59 | 73.00 | 5.11 |
| **Spectrum 4** | 18.27 | 4.55 | 72.41 | 4.76 |
| **Spectrum 5** | 21.42 | 4.32 | 69.44 | 4.83 |
| **Spectrum 6** | 21.23 | 4.52 | 69.17 | 5.07 |
| **Spectrum 7** | 21.21 | 4.52 | 69.25 | 5.02 |
| **Spectrum 8** | 21.39 | 4.53 | 68.96 | 5.12 |
| **Average** | 19.70 | 4.52 | 70.72 | 5.05 |
| **Normalized to Co** | 4.35 (39) | 1.00 (2) | 15.63 (38) | 1.12 (4) |

**$Mo_4RhGa_{16}Ge$**

| Spectrum # | Mo | Rh | Ga | Ge |
|---|---|---|---|---|
| **Spectrum 1** | 18.84 | 4.75 | 71.65 | 4.77 |
| **Spectrum 2** | 16.15 | 3.78 | 75.20 | 4.87 |
| **Spectrum 3** | 18.77 | 4.79 | 71.76 | 4.67 |
| **Spectrum 4** | 16.07 | 3.82 | 75.05 | 5.07 |
| **Spectrum 5** | 17.57 | 3.96 | 74.05 | 4.42 |
| **Spectrum 6** | 17.55 | 4.03 | 73.71 | 4.71 |
| **Spectrum 7** | 15.82 | 3.69 | 76.24 | 4.26 |
| **Spectrum 8** | 18.14 | 4.31 | 72.82 | 4.73 |
| **Average** | 17.36 | 4.14 | 73.81 | 4.69 |
| **Normalized to Rh** | 4.19 (29) | 1.00 (10) | 17.82 (40) | 1.13 (6) |

**$Mo_4IrGa_{16}Ge$**

| Spectrum # | Mo | Ir | Ga | Ge |
|---|---|---|---|---|
| **Spectrum 1** | 17.20 | 4.55 | 75.35 | 2.90 |
| **Spectrum 2** | 18.62 | 4.54 | 73.13 | 3.71 |
| **Spectrum 3** | 19.23 | 5.40 | 71.52 | 3.84 |
| **Spectrum 4** | 16.63 | 4.28 | 76.13 | 2.96 |
| **Average** | 17.92 | 4.69 | 74.03 | 3.35 |
| **Normalized to Ir** | 3.82 (26) | 1.00 (10) | 15.78 (45) | 0.71 (10) |

**Figure S3.** Temperature-dependent heat capacity ($C_p$) measured from 1.9 K to 40 K under 0 T external

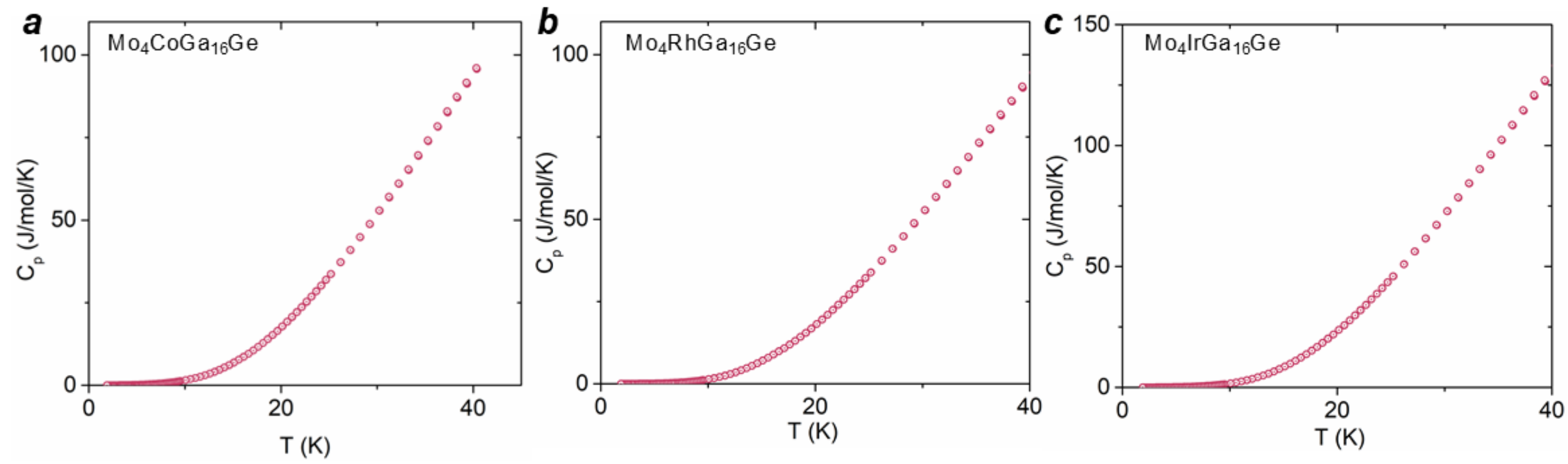

field for **a.** $Mo_4CoGa_{16}Ge$. **b.**$Mo_4RhGa_{16}Ge$ and **c.** $Mo_4IrGa_{16}Ge$.

**Table S4.** The fitted $\beta_1$ and $\beta_2$ value for $Mo_4TGa_{16}Ge$.

| **Parameters** | **$Mo_4CoGa_{16}Ge$** | **$Mo_4RhGa_{16}Ge$** | **$Mo_4IrGa_{16}Ge$** |
|---|---|---|---|
| **$\beta_1$ (mJ/mol/K$^4$)** | 0.391 (8) | 0.144 (7) | 0.574 (8) |
| **$\beta_2$ (mJ/mol/K$^6$)** | 0.014 (2) | 0.014 (1) | 0.009 (2) |